\documentclass[english,usenames,dvipsnames]{article}

\usepackage[T1]{fontenc}
\usepackage[latin9]{inputenc}

\usepackage{subfigure}
 
\usepackage{amsmath}
\usepackage{multirow}
\usepackage{tablefootnote}
\usepackage{comment}
\usepackage{epstopdf}

\newcommand{\beq}{\begin{equation}}
\newcommand{\eeq}{\end{equation}}
\newcommand{\bea}{\begin{eqnarray}}
\newcommand{\eea}{\end{eqnarray}}

\def\m1{M_1}
\def\m2{M_2}
\def\m3{M_3}

\def\br{\rm Br}

\def\ch10{\tilde \chi^0_1}

\def\mh{m_{h}}

\def\gev{\,{\rm GeV}}

\def\to{\rightarrow}

\newcommand{\lsim}{\mathrel{\mathop{\kern 0pt \rlap
  {\raise.2ex\hbox{$<$}}}
  \lower.9ex\hbox{\kern-.190em $\sim$}}}
\newcommand{\gsim}{\mathrel{\mathop{\kern 0pt \rlap
  {\raise.2ex\hbox{$>$}}}
  \lower.9ex\hbox{\kern-.190em $\sim$}}}
\newcommand{\mm}{\mu^+\mu^-}
\newcommand{\ee}{{e^{+} e^{-}}}

\newcommand{\mumu}{\mu^+\mu^-}
\newcommand{\ww}{W^+W^-}
\newcommand{\llll}{\ell^+\ell^-}

\newcommand{\s}{\sqrt{s}}
\newcommand{\shat}{  \sqrt{ \hat{s} } }
\newcommand{\h}{h}
\newcommand{\mgfive}{\textrm{MadGraph5\_aMC$@$NLO}}

\title{Higgs and BSM Physics \\at the future Muon Collider}

\author{Roberto Franceschini$^\dagger$ and Mario Greco$^\ddagger$ }

\newcommand{\lyxaddress}[1]{
	\par {\small \raggedright #1
	\vspace{0.04em}
	\noindent\par}
}
\usepackage{cite}
\usepackage{amstext}
\usepackage{graphicx}
\usepackage[unicode=true,pdfusetitle,
 bookmarks=true,bookmarksnumbered=false,bookmarksopen=false,
 linkcolor = ForestGreen,
            urlcolor  = NavyBlue,
            citecolor = CornflowerBlue,
 breaklinks=false,pdfborder={0 0 1},backref=false,colorlinks=true,linktocpage=true]
 {hyperref}
\date{}
\makeatletter
\begin{document}
 \maketitle

\lyxaddress{Dipartimento di Matematica e Fisica, Universit\`a degli Studi Roma Tre and 
 INFN, Sezione di Roma Tre, Via della Vasca Navale 84, I-00146 Rome, Italy\\
 \vspace{0.1cm}
 \hfill ${}^\dagger$ \href{roberto.franceschini@uniroma3.it}{roberto.franceschini@uniroma3.it}\; ${}^\ddagger$ \href{mario.greco@roma3.infn.it}{mario.greco@roma3.infn.it} \hfill }
 
  \abstract{We describe recent work on the physics of the Higgs boson at future muon colliders. Starting from the low energy muon collider at the Higgs boson pole we extend our discussion to the multi-TeV muon collider and outline the physics case for such machines about the properties of the Higgs boson and physics beyond the Standard Model that can be possibly discovered.\footnote{prepared for a \href{https://www.mdpi.com/journal/symmetry/special_issues/muon_collider}{ Special Issue {\it ``Physics Potential of the Muon Collider''}} of \href{https://www.mdpi.com/journal/symmetry/}{\it Symmetry}}\\

\tableofcontents

\newpage

\section{Introduction}

The opportunities offered by the realization of muonic beams have been realized long ago and the interest for this idea has been high for decades ~\cite{Budker:1969cd,Budker:1970ce,Skrinsky:1996um,Neuffer:1986dg,Blondel:387983
}. More recently, there has been more interest in the possibility of constructing a $\mm$ collider~\cite{ Ankenbrandt:1999cta,Geer:2009zz,Rubbia:2013iya,Palmer:2014nza,Boscolo:2018ytm} . Various surveys of the physics opportunities at such a collider have been made, see for example Refs. ~\cite{Barger:1996jm,Long:2020wfp}. It follows that a $\mm$ collider can essentially explore all the same physics that is accessible at an $e^+e^-$ collider of the same energy, but differently from the past, the time for a jump towards a future muon collider may now be finally ripe, as the possibilities for other more conventional types of colliders are shrinking and we are forced to think about bold and innovative new types of machines.  On the other hand the Higgs boson discovery at the LHC in 2012~\cite{Aad:2012tfa,Chatrchyan:2012ufa}  has opened a new era of particle physics and its properties absolutely need to be analysed with great precision and fully understood. The focus of any Higgs physics programme  is the question of how the Higgs boson couples to other Standard Model (SM) particles. Within the SM itself, all the couplings are uniquely determined, but possible new physics beyond the SM will modify these couplings in different ways, as the Higgs, for example, could be the portal to other gauge sectors. Then the Muon Collider (MC) opens up the particularly interesting possibility of direct $s$-channel Higgs production. In addition the MC is also a possible option for the next generation of high-energy collider machines, as it would allow achieving the highest energy frontier in lepton collisions, because muons do not suffer significant energy losses due to synchrotron radiation and therefore could be accelerated up to multi-TeV collision energies. 

Among many candidates of Higgs factories~\cite{Robson:2018enq,Abada:2019zxq,CEPCStudyGroup:2018ghi} the possibility of resonant production is especially important. 
The muon collider Higgs factory could produce the Higgs particle in the $s$-channel and perform an energy scan to map out the Higgs resonance line shape at a few MeV level. 
This approach would provide in principle the most direct measurement of the Higgs boson total width and the Yukawa coupling to the muons and other SM particles. However, 
 the extremely narrow width of the Higgs boson ($\Gamma$/M=$3.4\times 10^{-5}$) makes the resonant production rate very subject to any effect that shifts the collision center of mass  energy of the lepton collider.  Indeed there are two effects convoluted with the Higgs resonance production, the Beam Energy Spread (BES) and additional Initial State Radiation (ISR) corrections to the hard process which put severe limitations  to the observed production cross section ~\cite{Greco:2016izi}, by modifying the naive expectation of a sharp Breit-Wigner for  the Higgs resonance production.
 
The aim of the present paper is twofold. First we will review the Higgs $s$-channel resonance production, with a detailed discussion of the ISR effects, which play an important role in reducing the line shape cross section, and of the limitations from the BES in order to allow the appropriate precision of the experiments. In addition we will also discuss the background expectations on resonance for the main expected final states.

Besides  the possibility to make a Higgs boson factory using
muon beams, the possibility to store and accelerate large quantities
of muons opens the road to conceiving a high energy lepton collider
with circulating beams. In this work we will explore possible studies of the Higgs boson properties and associated new physics 
that are enabled by a high energy muon collider with center of mass energy in the multi-TeV regime.

The opportunities enabled by the availability of high energy bright muon beams are unique in the landscape of future colliders. In fact, most of the current projects of lepton collider operating at center
of mass energy well above the thresholds of SM states are linear colliders.
The reason is that electron and positron beams emit too large amounts
of synchrotron radiation if put in a circular orbit, therefore the
linear collider option is the only viable one if one wants to tame
synchrotron radiation. In order to reach multi-TeV center of mass
energy in linear colliders very innovative accelerator designs \cite{Robson:2018aa}
have been studied and tested in demonstrator facilities~\cite{Ruth:747483}.
Still, it seems hard to go beyond the 3 TeV center of mass energy
of the latest CLIC project stage. Despite the great amount of work
to optimize the innovative two-beam acceleration scheme of CLIC, it
remains very difficult to reach such large center of mass energy without
exceeding affordable amounts of wall-plug power requirements. Indeed
the 3 TeV stage of CLIC is estimated to require a yearly consumption
of electric energy in the range of few times the expenditure of the
future HL-LHC~\cite{1910.11775v2}. 

The power hungry character of linear electron-positron colliders is
not an isolated case in the landscape of particle colliders~\cite{Shiltsev:2016bfr,Shiltsev:2017aa}.
In fact, power requirements are a crucial bottleneck for the development
of $pp$ collider as well as lower energy $\ee$ colliders on circular
tunnels. For $\ee$ circular colliders this is obviously the consequence
of the synchrotron radiation we already mentioned. Even the most ambitious
programs under discussion, the CEPC~\cite{The-CEPC-Study-Group:2018ab,The-CEPC-Study-Group:2018aa}
and FCC-ee projects~\cite{Benedikt:2651299}, do not dare to consider
running above the $t\bar{t}$ threshold. Proton colliders also struggle
with synchrotron radiation as it is one of the main causes to heat
the superconducting magnets and significant power has to be put to
shield the magnets \cite{PhysRevAccelBeams.20.041002,1809.04330v1}
and keep them at the operating temperature. 

The possibility to circulate and handle muonic beams opens up a road
towards leptonic collision in the multi-TeV center of mass energy
ballpark, with manageable synchrotron radiation and affordable power
costs~\cite{Delahaye:2019aa,1808.01858v2}. Therefore a high energy muon collider
might be the pioneering project we need to set a new course for future
explorations in high energy physics. The jump in achievable center
of mass energy can be compared to the terrific progress that followed
milestones advances in particle accelerators such as the introduction
of beam-beam particle anti-particle collisions~\cite{Bernardini_1960,Cabibbo:1961sz,Bernardini:2004rp},
the use of superconducting materials in RF frequencies~\cite{Schmueser2002SuperconductivityIH}
or stochastic cooling for $p\bar{p}$ collisions~\cite{MARRINER200411}.

Then  we will examine the potential to explore new physics using these multi-TeV energy muon collisions using direct searches of new states, as well as indirect signals. Furthermore we will consider the possibility to stress test the SM using the copious production of SM states, e.g.   Higgs bosons and third generation quarks, measuring accurately properties of these SM states.

In section~\ref{lowE} we discuss the possibility of the $s$-channel resonant Higgs production and the parameterization of the BES and ISR effects. Then  we also discuss in detail their impact on the signal and the background for the main expected final states and on the global fits of Higgs properties. In section \ref{highE} we present the possibility to investigate Higgs physics and in particualr BSM physics in the Higgs sector at a multi-TeV muon collider. In this section we outline the several strategies that can be deployed at a multi-TeV muon collider thanks to the large momentum transfer available in reactions involving the beam particles and the low momentum ones from the ``partons'' within the scattering muons.
We present an  outlook and our conclusions in section~\ref{conclusions}.

\section{Low Energy Muon Collider \label{lowE}}
\subsection{Higgs boson resonant production}

The possibility of  $s$-channel resonant Higgs production is especially interesting ~\cite{Barger:1996jm} and indeed a muon collider can produce the Higgs boson resonantly at a reasonable rate. In the SM this measurement could probe the Higgs bosons width and the muon Yukawa directly. The clean environment of the lepton collider also enables precision measurement for many exclusive decays of the Higgs boson. This is particularly important also in the case of SUSY extensions of the SM, where the Higgs sector contains at least two Higgs doublets and the resulting spectrum of physical Higgs fields includes three neutral Higgs bosons, the CP-even $h_0$ and $H_0$ and the CP-odd $A_0$. The couplings of the MSSM Higgs bosons to fermions and vector bosons are determined by tan $\beta$ and the mixing angle  $\alpha$  between the neutral Higgs states $h_0$ and $H_0$. In addition all the Higgs bosons are produced in sufficient abundance in the $s$-channel muon-antimuon collisions to allow their detection for most of the parameter space. The Higgs boson widths are then crucial parameters, and for this study the muon collider is particularly suitable, also for providing tests of lepton universality of the Higgs couplings. 

From the accelerator's point of view, after the pioneering studies of the Muon Ionization Cooling Experiment (MICE) ~\cite{Bogomilov:2019kfj} , new suggestions have been recently put forward in order to realise a muon collider. First the proposal by C.~Rubbia ~\cite{Rubbia:2013iya,Rubbia:2019koh}, with a collider ring of radius of about 50m, which however also requires a powerful muon cooling process. Then, more recently, a low emittance muon accelerator (LEMMA)~\cite{Antonelli:2015nla,Amapane:2020det} has been suggested, using a positron beam on target, with the muons being produced in the electron-positron annihilation almost at rest, 
and the muon cooling is not necessary.

The extremely narrow width of the Higgs boson of about 4.1 MeV as predicted by the SM, makes the resonant production rate subject to any effect that shifts the collision c.m. energy of the lepton collider. Then there are two important effects related with the Higgs resonance production, the BES and the ISR corrections that make important modifications to the naive expectations. In a previous work \cite{Greco:2016izi} the convoluted effects of both BES and ISR have have been studied over the Breit-Wigner resonance for Higgs production at the muon collider. Their impact in different scenarios for both the Higgs signal and SM background has been considered. That study provides an improved analysis of the proposed future resonant Higgs factories and is also helpful for our understanding of the target accelerator design. In the following we will review those results, that have important implications for the experiments and also for the beam geometry design. 

Multiple soft photon radiation is an important effect which has to be taken into account when a narrow resonance is produced in the annihilation channel in lepton colliders. The first very clear example of such effect has been with the historical observation of J/Psi production in $e^+e^-$ annihilation  ~\cite{Augustin:1974xw} and
the origin  was soon discussed in very great detail~\cite{Greco:1975rm},  and also later for the case of the Z boson production~\cite{Greco:1980mh}. As a result a correction factor $\propto (\Gamma/M)^{4\alpha/\pi \log(2E/m)}$ modifies the lowest order cross section, where M and $\Gamma$ are the mass and width of the $s$-channel resonance, $W=2E$ is the total initial energy and m is the initial lepton mass. Physically this is understood by saying that the width provides a natural cut-off in damping the energy loss for radiation in the initial state. Very precise calculation techniques  for these QED effects have been developed for LEP experiments,  where in addition to multi-photon radiation finite corrections have been added, by including, at the least, up to two-loop effects, see for example Ref.~\cite{Nicrosini:1986sm}.
In the case of muon colliders, in particular for Higgs boson production studies,  those effects were not emphasized sufficiently in the past, and only recently their importance has been pointed out~\cite{Greco:2015yra,Jadach:2015cwa} for the experimental study of the Higgs line-shape as well as for the machine design of the initial BES. In particular the estimates of the reduction factors of the Higgs production cross sections, of order of 50\% or more, depending upon the machine energy spread, given in ref.~\cite{Greco:2015yra},  have been confirmed in ref.~\cite{Jadach:2015cwa}, and ref.~\cite{Greco:2016izi}, where the calculation techniques developed at the time of LEP experiments have been used, in order to estimate the expected precision of the theoretical results. 

Within the general formalism of the lepton structure functions, first introduced in Ref.~\cite{Kuraev:1985hb},  and later improved for LEP experiments,  defining  the probability distribution function $f^{\rm ISR}_{\ell\ell}(x)$ for the hard collision energy $x \sqrt {\hat s}$, then the hard collision cross section is written as
\beq
\sigma(\ell^+\ell^-\to h \to X)(\hat s)=\int d x~f_{\ell\ell}^{\rm ISR}(x;\hat s)\hat \sigma(\ell^+\ell^-\to h \to X)(x^2\hat s),
\eeq
where $x$ is the fraction of the c.~m.~energy at the hard collision with respect to the beam energy before the collision. Various approximations for the distribution function in the literature have been discussed in Ref. ~\cite{Greco:2016izi}, we will show the results obtained using the distribution function given in Ref.~\cite{Nicrosini:1986sm} which contains the full exponentiated term and the complete $O(\alpha)$  and $O(\alpha^2)$ terms.  

In addition, the observable cross section is given by the convolution of the energy distribution delivered by the collider. We assume that the lepton collider c.m.~energy ($\sqrt s$) has a flux $L$ distribution
\begin{equation}
\frac{dL(\sqrt{s})}{d\sqrt{\hat{s}}}=
\frac{ 1 }{ \sqrt{2 \pi \Delta } } \cdot \exp{ \left[ \frac{-( \sqrt{\hat s} - \sqrt s)^2} {2\Delta^2} \right] }\, , 
\end{equation}
with a Gaussian energy spread $\Delta  = R \sqrt {s}/\sqrt{2}$, where $R$ is the percentage beam energy resolution. Then the effective cross section is
\bea
\label{eq:convol}
&& \sigma_{\rm eff}(s) = \int d \sqrt{\hat s}\  \frac {dL(\sqrt s)} {d\sqrt{\hat s}} \  \sigma(\ell^+\ell^- \to h \to X)(\hat s) \\
\eea

For  $\Delta  \ll \Gamma_{h}$, the line shape of a Breit-Wigner resonance can be mapped out by scanning over the energy $\sqrt s$ as given in the first equation.
For $\Delta  \gg \Gamma_{h}$ on the other hand, the physical line shape is smeared out by the Gaussian distribution of the beam energy spread and the signal rate will be determined by the overlap of the Breit-Wigner and the luminosity distributions.

As a consequence of the ISR, a very significant phenomenon  is the ``radiative return'' to a lower mass resonance. Despite the beam collision energy is above a resonance mass, after ISR effects, the hard collision center of mass energy ``returns'' to the resonance mass and hits the Breit-Wigner enhancement again. This mechanism can be used to effectively produce lighter resonances without scanning the beam energy. 
On the other hand, when running at 125 GeV in a lepton collider the  amount of ``radiative return'' $Z$ bosons produced constitutes a large background for Higgs studies. 
One can easily see that different parameterizations of the ISR effects yield significantly different amount of ``radiative return'' $Z$ production rate. This  consideration clearly shows  the importance of a proper accurate treatment in evaluating the ISR effect.

\subsection{Numerical results on the ISR and BES on resonance}

The ISR effects, as discussed in the previous section, are very important and  need to be convoluted with the finite BES. We summarise numerically their combined effect in the Higgs boson production measurements in this section ~\cite{Greco:2016izi}. 

\begin{table}[tb]
\centering
\begin{tabular}{|c|c|c|c|c|}
  \hline 
$~{}~~~~~~~~\sigma$(BW) ${}~~~~~~$ & ISR alone & { R (\%)} &  BES alone &  BES+ISR  \\ \hline
\multirow{2}{*}{~71~{\bf pb}}\hfill & \multirow{2}{*}{37} & $0.01$ & $17$ & $10$\\ \cline{3-5}
 & & $0.003$ & $41$ & $22$ \\ \hline
%
\end{tabular}
\caption{Effective cross sections in units of pb  at the resonance $\sqrt {s}=\mh=125~\gev$, with Breit-Wigner resonance profile alone, with ISR alone, with BES alone for two choices of beam energy resolutions,
and both the BES and ISR effects included. 
}
\label{tab:muc_onpeak}
\end{table}

In Table~\ref{tab:muc_onpeak} we show the reduction effects for the resonance production of the SM Higgs boson at 125 GeV  including BES and ISR. 
The resonance production rate is reduced by a factor of about 2  with the inclusion of ISR effect. Independently, the production rate would be reduced by factors of 4.2 and 1.7 for beam spread of 0.01\% and 0.003\% respectively. The total reduction after the convolution of the beam spread and the ISR effect is 7.1 and 3.2 for the two beam spread scenarios, respectively.  A convenient analytical formula for evaluating the reduction factor on the peak as a function of the resonance width and  the machine energy spread, has been given in Ref.~\cite{Greco:2015yra}.

\begin{figure}[htbp]
\begin{center}
\includegraphics[width=0.48\textwidth]{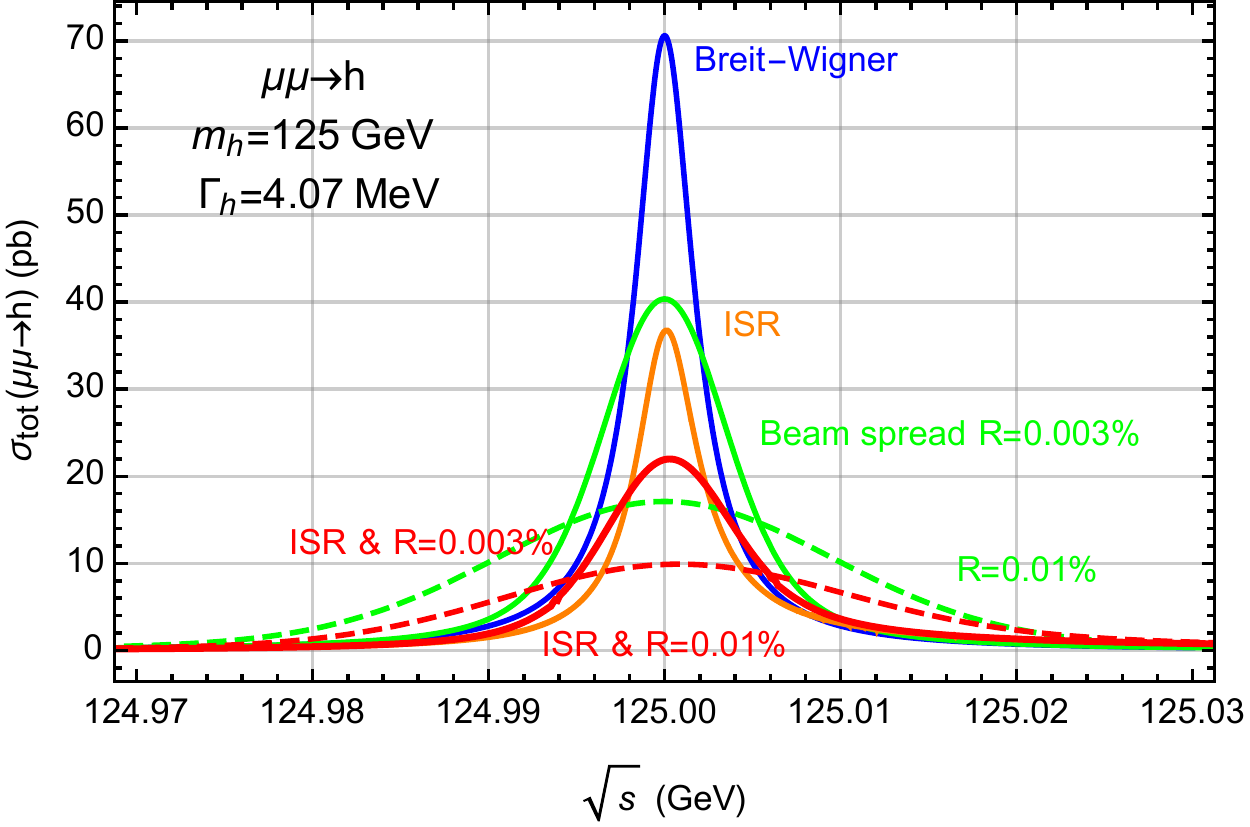}
\includegraphics[width=0.48\textwidth]{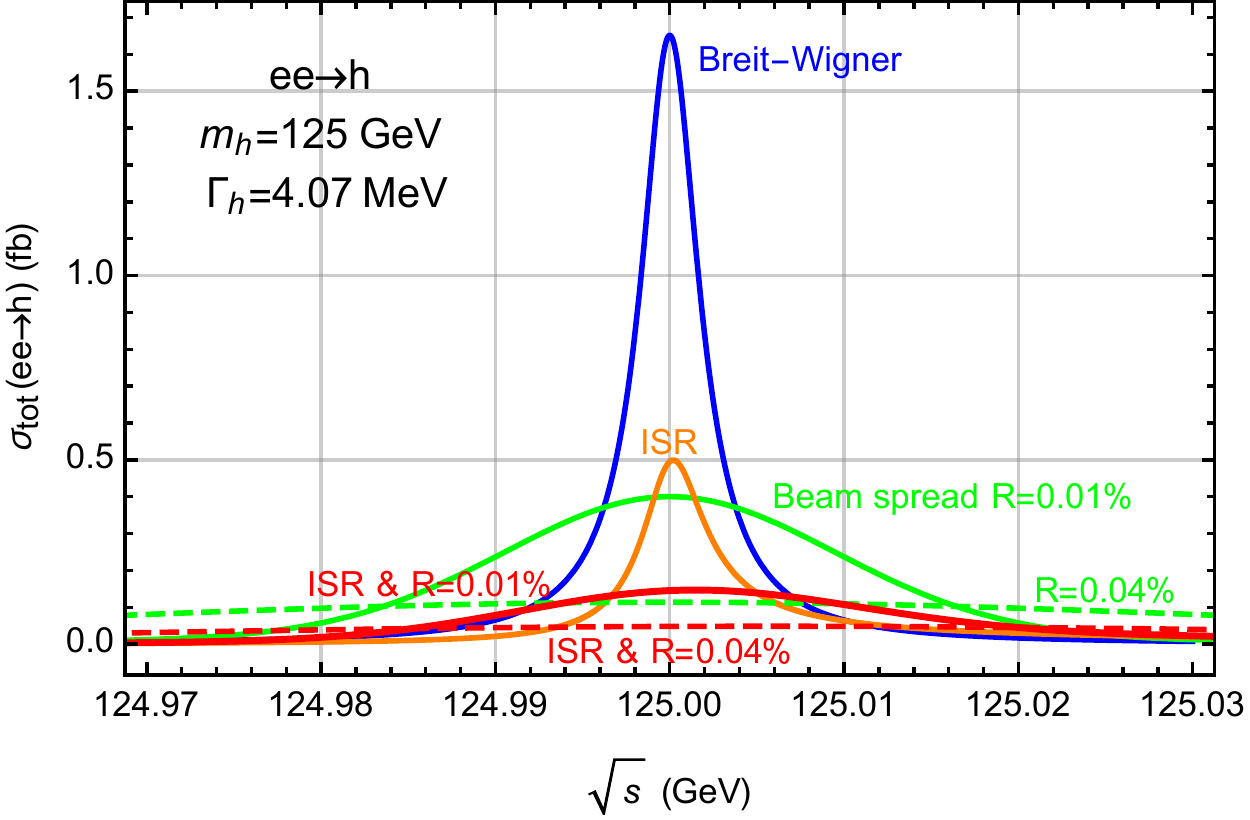}
\caption{
The line shapes of the resonances production of the SM Higgs boson as a function of the beam energy
$\sqrt{s}$ at a $\mm$ collider (left panel) and an 
$\ee$
 collider (right panel). The blue curve is the Breit-Wigner resonance line shape. The orange line shape includes the ISR effect alone. The green curves include the BES only with two different energy spreads. The red line shapes take into account all the Breit-Wigner resonance, ISR effect and BES  in solid and dashed lines, respectively. 
}
\label{fig:lineshape}
\end{center}
\end{figure}

\begin{figure}[t]
\centering
\includegraphics[width=0.48\textwidth]{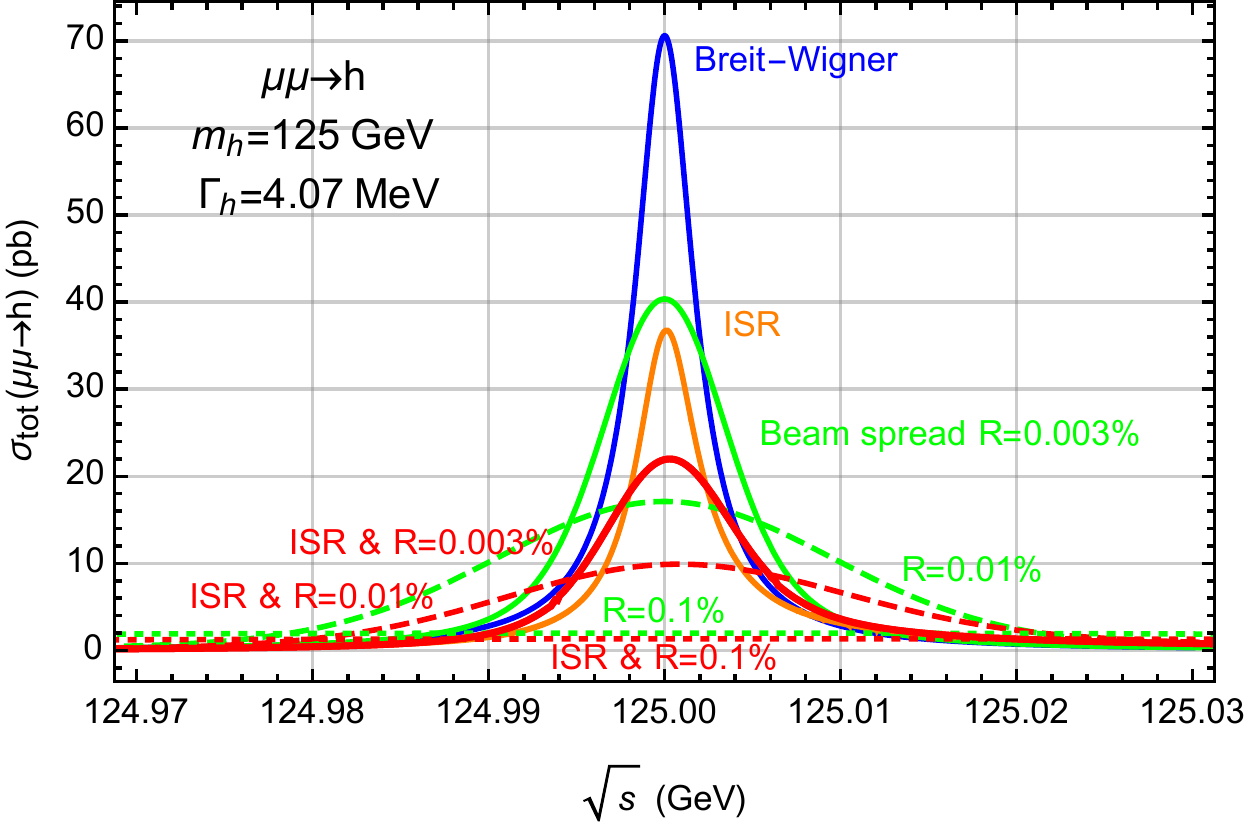} 
\caption{Same as in Fig.\ref{fig:lineshape}, with R =  $0.1\%$
\label{fig:Figure1}}
\end{figure}

The resulting line-shape from Ref.~\cite{Greco:2016izi}  is shown in Fig.~\ref{fig:lineshape} (left panel for the $\mm$ collider) for various setups of the parameters. For the reader's convenience we also show (right panel) the  ISR and BES results of Ref.~\cite{Greco:2016izi} for an electron-positron collider. The sharp Breit-Wigner resonance is in solid blue lines. The BES will broaden the resonance line-shape with a lower peak value and higher off-resonance cross sections, as illustrated by the green curves. The solid lines and dashed lines represent the narrow and wide BES of 0.003\% and 0.01\%, respectively.  In red lines we show the line shapes of the Higgs boson with both the BES and the ISR effect.  
The crucial role played by the numerical value of the BES parameter R is shown in Fig. 1.  When $R \gg\ 0.003\%$  the resonance signal is almost absent. This is clearly shown also in Fig.~\ref{fig:Figure1} for R = 0.1\%. The above analysis clearly indicates that a muon collider resonant Higgs factory makes sense only if the initial beams energy spread is of order of the Higgs width.

\begin{table}[tb]
\centering
\begin{tabular}{|c|c|c|c|c|c|}
  \hline
 & $\mm\rightarrow h$ & \multicolumn{2}{|c|}{$h\rightarrow b\bar b$} &  \multicolumn{2}{|c|}{$h\rightarrow WW^*$}  \\ \cline{3-6}
 \raisebox{1.6ex}[0pt]{R (\%)} & $\sigma_{\rm eff}$ ({\bf pb}) & $\sigma_{Sig}$ & $\sigma_{Bkg}$ &  $\sigma_{Sig}$ & $\sigma_{Bkg}$ \\ \hline
 $0.01$ & $10$ & $5.6$ & & $2.1$ &  \\ \cline{1-3} \cline{5-5}
$0.003$ & $22$ & $12$ & \raisebox{1.6ex}[0pt]{$20$}  & $4.6$ & \raisebox{1.6ex}[0pt]{$0.051$}\\ \hline
\end{tabular}
\caption{Signal and background effective cross sections at the resonance $\sqrt {s}=\mh=125~\gev$  in pb, for two choices of beam energy resolutions $R$ and two leading decay channels with ISR effects taken into account, with the SM branching fractions $\br_{b\bar b}=58\%$ and $\br_{WW^*}=21\%$. For the $b\bar b$ background, a conservative cut on the $b\bar b$ invariant mass to be greater than 100 GeV is applied.
}
\label{tab:muon_sigbkg}
\end{table}

In addition to the Higgs signal, an important issue of phenomenological interest is the question of the expected background in the various Higgs decay channels. This is related to the tail of the Z-boson production in the lepton annihilation. This issue has been discussed in detail in Ref. ~\cite{Greco:2016izi}. We will report here the main conclusions. The main search channels will be the  exclusive mode of $b\bar b$ and $WW^*$.  For the $b\bar b$  final state the main background is from the off-shell $Z/\gamma$ $s$-channel production.
The ISR effect does increase the on-shell process  $Z\to b\bar b$ background through the radiative return by a factor of seven. This can be reduced by imposing an invariant mass cut of about 100 GeV which leads to around 20\% increase in such background comparing to the tree level estimate. Alternatively one can foresee a cut on the angle between the two b-jets, which could be measured more precisely than the invariant mass. 

Beyond the $b\bar b$ final state, another major channel for muon collider Higgs physics is the $WW^*$ channel. In this channel the  background from the SM process is quite small.  The ISR effect introduces no ``radiative return'' for such process. Consequently, the background rate does not change from the tree-level estimate.  We summarize in Table~\ref{tab:muon_sigbkg} the on-shell Higgs production rate and background rate in these two leading channels with the inclusion of the ISR and BES effects. We can see from the table that at the muon collider Higgs factory, the signal to background ratio is pretty large and the observability is simply dominated by the statistics.

\subsection{Outlook on low energy options}
We have discussed  the  $s$-channel resonant Higgs production in a future muon collider together with  the effects from the initial state radiation and the beam energy spread.
We have quantified their impact for  different representative choices of the BES for both the Higgs signal in its main decay modes and the corresponding SM background. 
We have shown  that
the BES effect is potentially the leading factor for the resonant signal identification, and it alone reduces the on resonance Higgs production cross section by a factor of 1.7 (4.2) for a muon collider with $R=0.003\%$ ($R=0.01\%$).
Then the ISR effect alone reduces the on-resonance Higgs production cross section by a factor of about 2. 
 The total reduction factors for the on-resonance Higgs production cross section after convoluting the BES and ISR effects are 3.2 (7.1) for a muon collider with $R=0.003\%$ ($R=0.01\%$). 
Therefore the BES parameter R plays a cruciale role and the above analysis clearly indicates that a muon collider resonant Higgs factory makes sense only if the initial beams energy spread is of order of the Higgs width.
In addition the background for the $h \to b\bar b$ channel is increased by a factor of  seven due to the ``radiative return'' of the $Z$ boson  and a cut on the minimal $b\bar b$ invariant mass of 100 GeV reduces such background, resulting in an increase of the tree-level estimate of the background by 20\%. Then both the $h\to b\bar b$ and $h\to WW^*$ contribute to the signal sensitivity. 

Significant efforts are needed to achieve these ambitious targets on the beam parameters for a muon collider Higgs boson factory at the Higgs boson pole. Given the size of the effort needed it is important to carefully gauge the reward that one could obtain in the knowledge of the Higgs boson from such Higgs boson factory. Considering realistic integrated luminosities O(fb$^{-1}$) the Higgs pole muon collider would produce at best a fraction of $10^5$ Higgs bosons. Statistical uncertainties from such a data set  can lead to couplings determinations in the ballpark of 1\% precision for the most abundant Higgs boson decay channels, thus potentially improving on the most optimistic Higgs boson couplings determination that the HL-LHC could give \cite{Blas:2019aa}.

The Higgs pole muon collider Higgs boson factory would also have the great advantage over the HL-LHC to be able to measure the Higgs boson width directly~\cite{Han:2013aa,1304.5270v1,Greco:2016izi,JanotFacultyCERN2018,LiuPITT2020}. Such a measurement is one of the few possible ways to obtain absolute measurements on the Higgs couplings, therefore would be a cornerstone for our knowledge of the Higgs boson and will have impact on any future study of the Higgs boson. As a matter of fact, even in presence of a large data sets, e.g. the  O($10^6$) $Zh$ pairs produced at the circular $e^+e^-$ machines at the $Zh$ threshold considered in Refs.~\cite{1810.09037v2,Benedikt:2653669,Mangano:2651294}, the best known quantities will be dimensionless ratios, e.g. ratios of rates or ratios of branching fractions, whereas we will have significantly worse knowledge of the overall scale of these rates and absolute Higgs couplings. Remarkably, the knowledge of an absolute coupling scale from the Higgs boson width measurement that can be carried out at a muon collider Higgs boson pole factory could come quite close to what is doable by measuring the total $Zh$ rate and using the recoil method in hadronic $Z$ and leptonic $Z$ boson events, with no requirements on the decay of the Higgs boson at $e^+e^-$ factories operating at 240 GeV and above. The actual result on the Higgs boson width and its impact on the overall determination of the Higgs boson couplings is an active subject of study~\cite{MuCHiggsPaper}. Preliminary results \cite{MuCHiggsPaper,LiuPITT2020} indicate that a clean extraction of the Higgs boson width can lead to couplings determinations that may be only slightly inferior to the performance of a Higgs boson factory at the $Zh$ threshold \cite{Benedikt:2653669,Mangano:2651294,1810.09037v2} or at higher energies \cite{Abramowicz:2016zbo}.

\section{High energy muon collider\label{highE}}

\subsection{First of a new kind}

The idea of a high energy muon collider has been put forward since decades~\cite{Budker:1969cd,Budker:1970ce,Skrinsky:1996um,Neuffer:1986dg,Blondel:387983} and the possibility to use muon beams to go to the highest energies is definitively not new.
Differently from the past, the time for a jump towards a future muon
collider may now be finally ripe, as the possibilities for other more
conventional types of colliders are shrinking and we are forced to
think about bold and innovative new types of machines. 

In fact, future electron positron circular machines and $pp$ colliders
are essentially based on the same type of technology that enabled
the Large Hadron Collider (LHC) and the Large Electron Positron collider
(LEP). Of course, improvements have been possible over the years of
operation of these types of machines and will still be possible in
the future application of these technologies~\cite{HLlhcESUinput,Ohnishi:2013fma},
but it is  fair to say that  presently discussed future $\ee$
and $pp$ colliders~\cite{Mangano:2651294} are mostly bigger and
not fundamentally different from their predecessors. 

The developments necessary to build these bigger $\ee$ and $pp$
machines, e.g. in superconductors technology~\cite{Rossi2015TheEF,Gourlay2018SuperconductingAM},
represent great challenges and might have enormous societal and technological
impact. These advances are definitely worth a strong R\&D program,
as indicated by the recent update of the European Strategy for Particle
Physics~\cite{CERN-ESU-014,CERN-ESU-013}. Still, these futuristic
$\ee$ and $pp$ machines will be ``just'' more powerful versions
of machines we have already built. 

On the contrary, a high energy muon collider will be the first of
a new kind of machines and will open the way to a novel investigation
of fundamental interactions at the shortest distances, significantly
improving over the physics capabilities of more traditional machines
in most kind of investigations. The recent surge of phenomenological studies on the physics case of the muon collider \cite{2103.14043v1,2103.12766v1,2103.09844v1,4topsMuon,2103.01617v1,2102.11292v1,2102.08386v1,2101.10469v1,2101.10334v1,2101.04956v1,2012.11555v1,2012.02769v1,2010.05915v1,2011.13949v1,2011.03055v1,2008.12204v1,2009.11287v1,2007.14300v2,2006.16277v1,2005.10289v1,2003.13628v1,2001.04431v1,Ruhdorfer:2019ab,Di-Luzio:2018aa,Buttazzo:2018aa,Fornal:2018ab,Chakrabarty:2015rf} 
is a proof of the enormous interest on this machine.

In the following we will
outline the tracks along which a high energy muon collider can investigate
new physics at the energy frontier. We will highlight the strengths
of the investigations enabled by high energy muon collisions and we
will also highlight crucial requirements that need to be met by this
kind of machine or risk to jeopardize the outcome of these investigations. 

\subsection{Multiplexing the search for new physics}

A distinctive feature of high energy muon colliders in searching for
new physics is that they can operate different search modes and it
is possible to obtain very strong bounds from different types of searches.

For a quick categorization of search modes that can be pursued at
a high energy muon collider we can divide searches in:
\begin{itemize}
\item Direct production of new physics, e.g. the on-shell production of
new states $\llll\to\chi\chi$ where $\chi$ is a new physics particles,
for instance a dark matter particle;
\item Indirect effects from off-shell new physics, e.g. the modification
to the angular distribution of $\llll\to f\bar{f}$ Drell-Yan processes
due to contact interactions $\bar{\psi}_{\ell}\psi_{\ell}\bar{\psi}_{f}\psi_{f}$
beyond the SM;
\item Copious production of SM states in (effective) $2\to1$ annihilations,
$2\to2$ scatterings, or 3-body and multi-body productions. These
include for instance the effective $W$ boson annihilation to produce
Higgs bosons in $\llll\to\nu\nu\h$, $\llll\to t\bar{t}$ and $\llll\to t\bar{t}\h$
processes.
\end{itemize}
In all these search modes a high energy muon collider will result
in significant advances compared to the HL-LHC and, in most cases,
even in comparison to proposed ambitious future collider projects. 

We will briefly discuss examples of these search strategies in the
following. For now we can highlight that the key feature of a high
energy muon collider that enables all these search strategies is the
possibility to have \emph{both} a large center of mass energy \emph{and}
at the same time keep a relatively clean collision environment. A
high energy muon collider can operate as a clean lepton machine and
at the same time have reach over the energy frontier comparable, if
not superior, to hadronic machines.

Furthermore a high energy muon collisions has the great operational
advantage that the searches outlined above can be pursued \emph{at
the same time}, without requiring dedicated runs or machine settings.
Due to the largely unknown character of new physics this fact is very
important, as it implies that the operation of a high energy muon
collider does not require to commit to one strategy ahead of time
or to make hard choices in allocating machine run time or planning
stages of its construction.

\subsection{Direct production of new physics}

Muons, being point-like particles, have the great advantage to make
all of their energy available to produce heavy final states. This
needs to be contrasted with protons, for which we are forced to talk
about partons and energy fractions carried by them from the very start
of description of the collisions.

\begin{figure}
\begin{centering}
\includegraphics[width=0.67\linewidth]{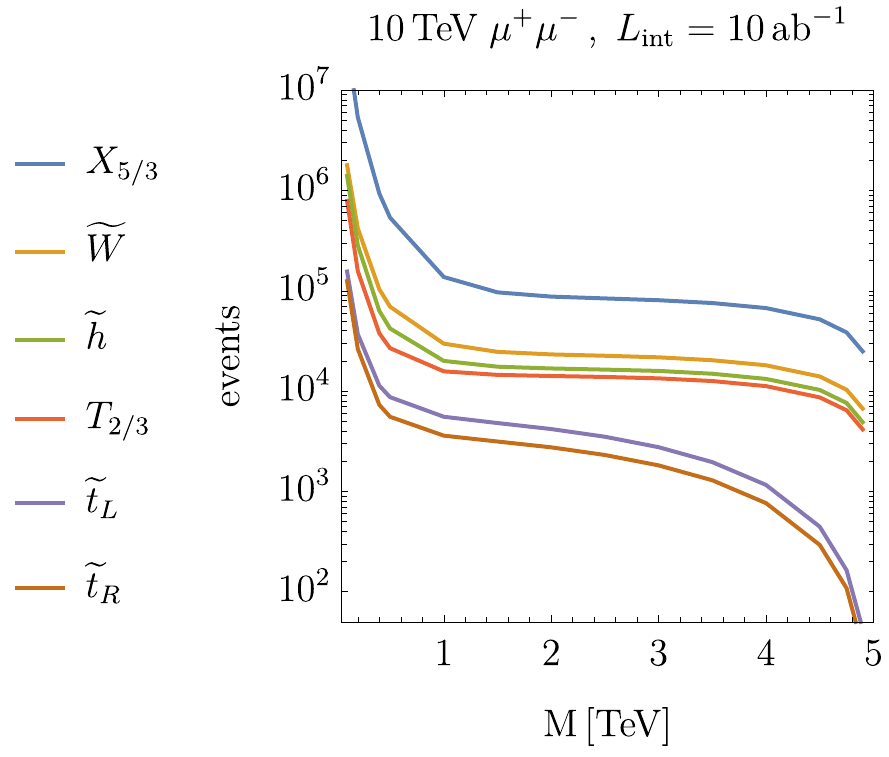}
\par\end{centering}
\caption{\label{fig:Rates-for-direct-BSM}Rates for direct production of new
states. The labels follow standard nomenclature of composite Higgs models and supersymmetric models. However we computed cross-sections using only gauge interactions, whereas in these models each state may have specific model dependent interactions that can increase their production rate. Therefore our labels are shorthands for $SU(3) \otimes SU(2)\otimes U(1)$ charges: fermions ${X}_{5/3}\sim (3,2)_{7/6}$, $\tilde{W}\sim (1,3)_0$, $\tilde{h}\sim(1,2)_{\pm1/2}$, ${T}_{2/3}\sim(3,1)_{2/3}$ and scalars $\tilde{t}_L \sim (3,2)_{1/6}$, $\tilde{t}_R \sim (3,1)_{2/3}$.} 
\end{figure}

As muons carry electric and weak gauge charges they are excellent
initial states to produce any state, SM or BSM, that has electric
or weak gauge charge. Assuming only these gauge interactions for new
physics states, and barring any non-gauge interactions for the moment,
we can compute cross-section for the production of heavy states at
a high energy muon collider. They are reported in Figure~\ref{fig:Rates-for-direct-BSM}
as number of event at a 10 TeV collider for 10/ab integrated luminosity.
The figure is taken from Ref.~\cite{2012.11555v1} and displays the
result for a set of new physics states using a supersymmetric nomenclature,
but without having in mind any supersymmetric model of sort. The name
$\tilde{t}_{L}$ should just be read as a shorthand for the $SU(3)\otimes SU(2)\otimes U(1)$
gauge quantum numbers, that in this case are $(3,2)_{1/6}$ . We stress
that only gauge interactions are considered in this result.

The reported cross-section contain two main contributions: \emph{i)}
the direct production in $2\to2$ Drell-Yan process; \emph{ii)} the
production from gauge boson fusion from the flux of equivalent vector
bosons that is part of the muon beam quantum structure, i.e. it follows
from the $\mu\to W\nu$ or $\mu\to\gamma\mu$ and $\mu\to Z\mu$ splittings.

The Drell-Yan production is essentially given by the gauge couplings
and the geometrical factors of the cross-section, so that for a particle
with couplings of order $O(1)$ we expect a $2\to2$ cross-section
\begin{equation}
\sigma\simeq O(1)\textrm{ fb} \cdot\left(\frac{10\textrm{ TeV}}{\s}\right)^{2} \,.\label{eq:generic-rate}
\end{equation}
\\
This cross-section may be larger in case of large multiplicities in
the final state due to spin or color quantum numbers, hence scalars
have smaller cross-sections than fermions and colored particles have
larger particles than particles without color.

\begin{figure}
\begin{centering}
\includegraphics[width=0.67\linewidth]{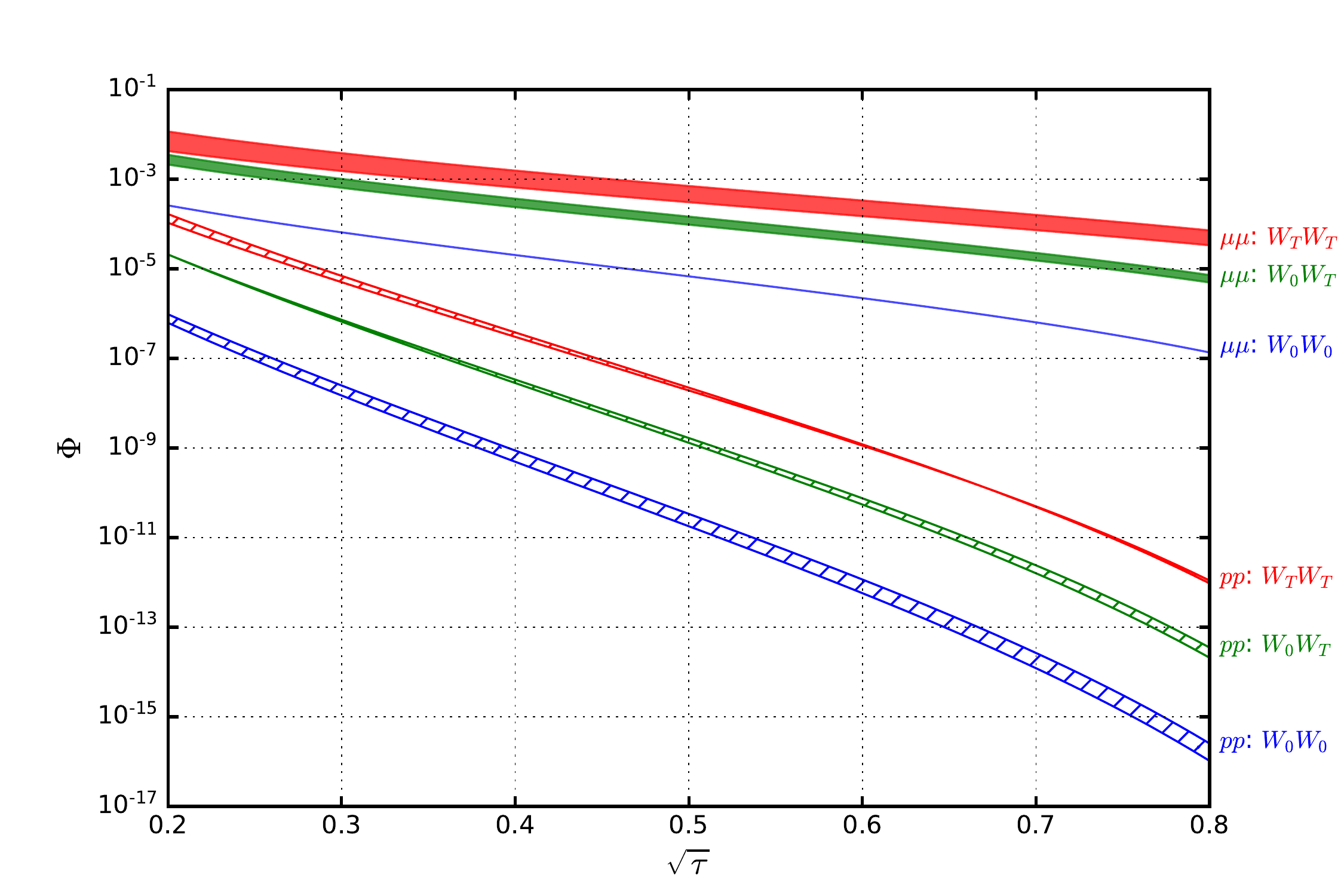}
\par\end{centering}
\caption{\label{fig:Partonic-fluxes}Partonic flux for transverse-transverse
(red), transverse-longitudinal (green), longitudinal-longitudinal
(blue) $W$ boson in $pp$ (hatched) and $\mu\mu$ (solid shading)
collisions.}
\end{figure}

When the mass of the produced particle is light compared to the center
of mass energy of the collider it is possible to efficiently produce
particles from collisions of equivalent bosons, e.g. 
\[
WW\to\chi\chi\,,
\]
from $W$ boson radiated off the beams. In the collinear radiation
approximation these collisions can be factorized and we can talk about
an effective $W$ boson beam and effective $W$ bosons fusion. This
process suffers the decrease of the partonic luminosity of the $W$
bosons at large $WW$ center of mass energy. The flux of possible
polarization states of $W$ partons in the muon beam structure can
be seen in Figure~\ref{fig:Partonic-fluxes} from Ref.\cite{2005.10289v1}
and it roughly falls a fifth power of the $WW$ center of mass energy
tracker variable $\sqrt{\tau}=m_{WW}/\s$.

Clearly the boson fusion processes can give significant enhancements
of the total rates for new physics states, but are largely subdominant
to $2\to2$ Drell-Yan close to the kinematic reach of the machine.
Therefore the reach over the energy frontier of a high energy muon
collider can be estimated simply looking at the Drell-Yan rates. With
this prescription in mind, it is clear that the mass reach of a muon
collider for new physics states can be quite close to $\s/2$. The
actual reach depends on how spectacular or subtle is the decay mode
of the newly produced particles, but the relatively clean collision
environment puts us in favorable conditions to go after even somewhat
subtle signatures. This should be contrasted with the typical situation
of hadronic colliders in which it is relatively easy to hide copiously
produced particles by just having them decay into soft enough final
states that can be easily produced in SM reactions.

An educated guess for a model-independent comparison of $pp$ and
$\llll$ direct reach for new physics can be found in Ref.\cite{2005.10289v1}.
The outcome is that a 14 TeV center of mass energy $\llll$ collider
can be as powerful to probe directly heavy new physics as a 100 TeV
$pp$ collider.
Concrete studies for the production of heavy Higgs bosons weak doublets or singlets confirm an excellent sensitivity to the production of these states in vector boson fusion for singlets~\cite{Buttazzo:2018aa} and in pair production via gauge interactions or other associated productions for doublets~\cite{2102.08386v1}. 

It is important to stress that because of the expected cross-section
for an electroweak Drell-Yan process in eq.(\ref{eq:generic-rate}),
the luminosity requirement for a discovery with a O(10) new physics
events is a rather low integrated luminosity of order O($10^{-2}$)~$\text{ab}^{-1}$
for a 10~TeV machine. This is considerably below what has been found
would be doable in the MAP study~\cite{Palmer_2014} for a proton-sourced
muon collider and is largely below the luminosities consider even
in ``luminosity-hungry'' linear $\llll$ colliders~\cite{Robson:2018ab}.
In conclusion the direct search for new physics is major driver towards
increasing the attainable $\s$ , but does not pose serious constraints
for what concerns the luminosity of these machines.

\subsection{Indirect effects from off-shell new physics}

Indirectly testing new physics requires studying well measurable quantities
for which precise and reliable predictions are available from theory.
A classic example for leptonic colliders are the angular distributions
of final states, which can reveal heavy new physics beyond the kinematic
reach of the machine, e.g. a hint of heavy weak bosons from processes
dominated by QED~\cite{Brandelik:1982id}.

While the accuracy of theoretical calculations available might change
from now to the time of operation a high energy muon collider, the
possibility to measure accurately simple quantities such as total
or fiducial rates can be largely anticipated today by just looking
at the statistical uncertainties expected for the rates of interests.
For a process with cross-section $\sigma$ we expect to collect a
number of events $N=\sigma\cdot\mathcal{L}$, where $\mathcal{L}$
is the total luminosity collected at the experiment. As we are considering
high energy processes we can roughly estimate cross-sections by dimensional
analysis $\sigma\sim1/s$, where $s$ is the characteristic energy
scale of momentum transferred in the scattering and the exact coefficients
are determined by the coupling constants of phase-space factors of
each process. In particular for a $2\to2$ scattering we gave the
estimate eq.(\ref{eq:generic-rate}), which is valid for the production
of SM as well as BSM states. According to this estimate, if we take
a luminosity

\begin{equation}
\mathcal{L}=10\text{ ab}^{-1}\cdot\left(\frac{\s}{10\text{ TeV}}\right)^{2}\label{eq:baseline-lumi}
\end{equation}
we obtain 
\[
N=\sigma\cdot\mathcal{L}=10^{4}\text{ events independently of }\sqrt{s}\,.
\]
This number of events is apt to carry out precision measurements at
the 1\% level. Of course the exact number of events usable in the
measurement and the exact meaning of ``precision'' needs to be qualified
further, but we find this estimate nevertheless useful as it poses
a rough target for any plan to use a muon collider to carry out precision
studies. Indeed, obtaining a smaller number of events would result
in measurements at the O(10\%) level which can hardly be called ``precise''. 

To put these considerations on firm ground we need to compute the
expected size of the effects from new physics. Taking into account
that a machine running at center of mass energy $\sqrt{s}$ can directly
produce and discover new particles with mass below $\sqrt{s}/2$ it
makes sense to consider effects from new physics heavier, and possibly
much heavier, than the direct production limit $\s/2$. The effects
of these state of mass $M\gg\s$ can be encapsulated in a number of
new interactions vertexes that are contact interactions among SM states,
e.g. a four-fermion contact interaction such as that of the Fermi
theory of weak interactions at energies well below the mass of the
$W$ boson. When we study reactions in which new contact interactions
can mediate the scattering we obtain contributions to the scattering
amplitude weighed by powers of 
\begin{equation}
\epsilon=g^{2}\left(s/M^{2}\right)\,.\label{eq:expansion-parameter}
\end{equation}
As a result, the interference between SM and BSM sub-amplitudes contributes
to the cross-section with an $\s$-independent term, which can cause
measurable deviations in sensitive observables. The size of these
deviations with respect to the SM is controlled by $g^{2}s/M^{2}$
and it cannot be larger than a fraction of $(16\pi^{2})s/M^{2}$,
therefore we expect small effects from new physics of mass $M\gg\s$.
Indeed if these effects were big, we would have already got hints
at the LHC and direct production of new states would be a more suitable
way to search for new physics at the muon collider. Having in mind
the ``unit of measurement'' in eq.(\ref{eq:expansion-parameter})
we understand why 1\% to 10\% starts being an interesting level of
precision to probe new physics indirectly. This is about the largest
effect one can expect for new physics heavy enough to escape the direct
production and be well within the approximation we make by taking
perturbative values of $g$ in the expansion parameter of our EFT
in eq.(\ref{eq:expansion-parameter}).

\subsubsection{The size of the Higgs boson}

To be concrete we want to discuss the example of new physics indirect
effects in precision studies of quantities related to the Higgs boson,
and in particular to both the physical Higgs boson and the would-be
Goldstone bosons eaten in the massive gauge vector bosons in the Higgs
mechanism. This example highlights extremely well the power of a high
energy muon collider to study the Higgs sector of the Standard Model. 

The theoretical setting in which we can carry out this study is a
dimension-6 EFT that extends the SM, the so-called SMEFT~\cite{WarsawBasis:2010es},
or an equivalent EFT in which the Higgs boson is more directly involved
in the new contact interactions. While these ``bases'' for the EFT
can be shown to be equivalent in physical results, the SILH basis~\cite{Giudice:2007yg}
is more transparent for our study as it highlights the effects on
Higgs, Goldstone and gauge bosons interactions.

The dim-6 lagrangian that extends the SM in the SILH basis has a large
numbers of terms. Dealing with them all at once requires to carry
out a large number of measurements to constrain each contact interaction
using a sensitive measurement. To reduce the complexity of this task
we can approach the problem with some theoretical picture that provides
rough estimates on the size of each of the many contact interactions.
This amounts to image a concrete dynamics for the UV lagrangian that
gives rise to the low energy EFT and provides us with a rough parametrization
of the size of each contact interactions. The size of the contact
interaction couplings can be expressed in terms of powers of the fundamental
parameters of the UV lagrangian, up to numerical factors that depend
on the specific UV lagrangian and that are not interesting to obtain
just an estimate of the size of the BSM interactions. This is the
so-called ``power counting'' which allows to estimate the size of
interactions strength from a generic type of UV completion of the
EFT.

In the SILH case a power counting for generic UV completion describes
the possibility that the Higgs boson is not a point-like particle,
but it has a finite size $\ell_{H}\sim1/m_{\star}$ which is about
the order of magnitude of the mass of heavy new physics that belongs
to the UV completion of the low energy SILH EFT. In this picture the
Higgs boson is a light particle of the theory valid at and above the
EFT scale and it just happens to be light enough for us to produce
it and study it. 

The position of the Higgs bosons is then similar to the position of
the pions in the world of hadrons. They are light enough that one
can study pion scattering (or Higgs boson physics) even if the available
energy is limited below the mass of the first $\rho$ meson and the
other heavier hadrons. Pions are therefore both part of a low-energy
EFT, the so called sigma model of pions, but are also the first of
a long list of hadrons, which eventually can be understood all as
low energy manifestations of more fundamental quarks and gluons. The
lightness of the pions can be understood from the fact that they are
Goldstone bosons of an underlying symmetry of QCD, broken by the quark
masses. Thus, at least when the symmetry is not too badly broken,
pions and other kind of Goldstone bosons are expected to be lighter
than other states. 

The picture of the SILH is to imagine the Higgs boson is a light composite
particle, like the pions, that emerges as pseudo-Nambu-Goldstone boson
of the symmetry breaking pattern of the UV completion of our low energy
EFT. The study of precision observables involving the Higgs bosons
and the eaten Goldstone bosons of the SM can be seen as the study
of pions in search for the evidence of indirect effects of heavier
$\rho$ mesons. The mass of said $\rho$ meson can be seen as the
energy scale of momentum transferred at which we start probing distances
so short that the structure of the Higgs boson starts to emerge, fully
displaying its finite size. 

\begin{figure}
\centering{}\includegraphics[width=0.67\linewidth]{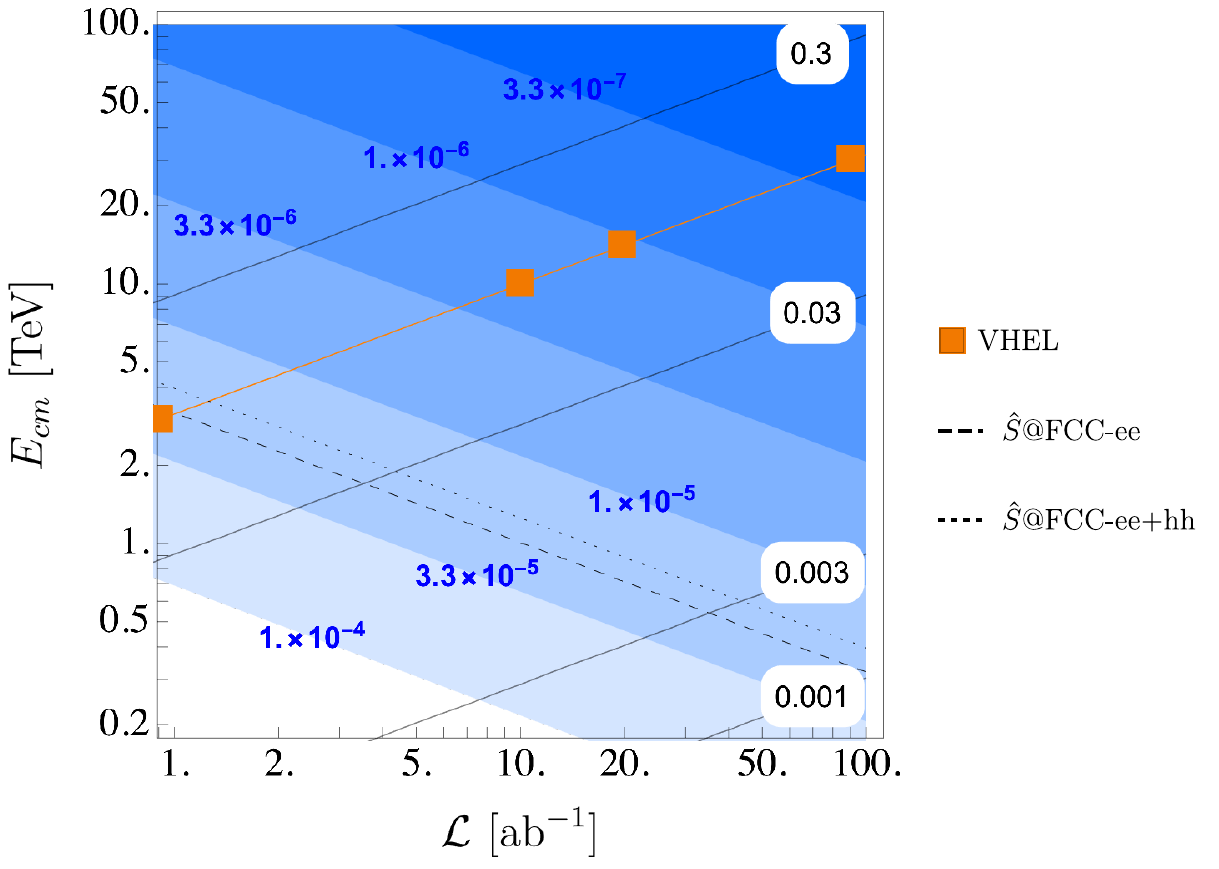}\caption{\label{fig:Lower-bounds-on-mStar}Lower-bounds on $m_{\star}=1/\ell_{H}$
expressed as upper bounds on $\hat{S}=(m_{W}/m_{\star})^{2}$ (blue
shades and labels). Dashed lines corresponds to limits on the same
quantity from the combination of $ee$ Higgs factory and high energy
$pp$ colliders $FCC{\rm -ee}$ an $FCC{\rm -hh}$~\cite{1910.11775v2}.}
\end{figure}

\begin{figure}
\begin{centering}
\includegraphics[width=0.67\linewidth]{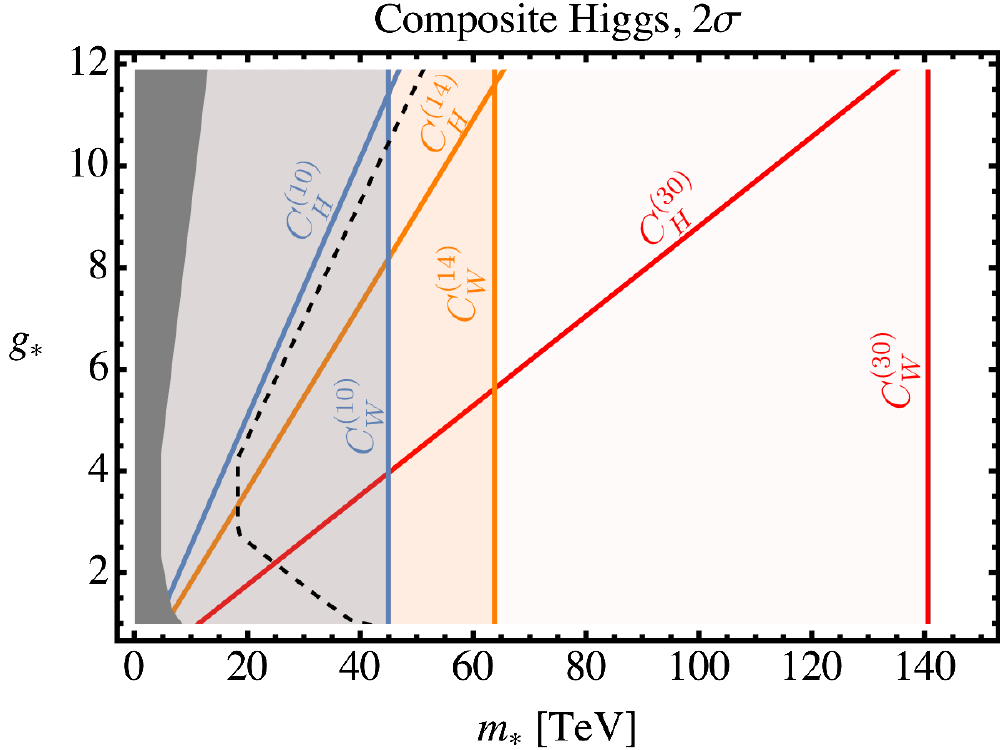}
\par\end{centering}
\caption{\label{fig:Bounds-Mstar}Bounds from Ref.~\cite{2012.11555v1} on
the size of the Higgs bosons $\ell_{H}\simeq1/m_{\star}$ from a 10~TeV
(blue), 14~TeV (orange), 30~TeV (red) $\mumu$ collider using the
luminosity eq.(\ref{eq:baseline-lumi}). The vertical lines are from
di-boson and multi-boson production (e.g. $\ww,Z\h,\ww\h$). Diagonal
lines are from $hh$ production. Bounds con $m_{\star}$ depend on
a generic coupling $g_{\star}$ as suggested by the SILH power counting.
The dashed line corresponds to the limits projected for the CLIC 3
TeV stage~\cite{1910.11775v2}. The solid shade corresponds to the
bounds from HL-LHC~\cite{1910.11775v2}.}
\end{figure}

In order to probe the size of the Higgs boson a very effective strategy
consists in studying Higgs and Goldstone bosons scattering in all
possible production processes. The effects of the finite size of the
Higgs boson are enhanced by the momentum transferred in the reactions,
therefore this study has a clear demand for high energy. Nevertheless,
a number of collisions to measure rates with sufficient precision
is necessary to put bounds. In Ref.~\cite{2012.11555v1} the number
of events for the reaction $Z\h$ expected for luminosity eq.(\ref{eq:baseline-lumi})
has been translated into an expected bound on the size of the Higgs
boson for generic $\s$ muon collider. This bound corresponds to the
orange line in Figure~\ref{fig:Lower-bounds-on-mStar} on which we
highlighted center of mass energies 3~TeV, 10~TeV and two ``10+ TeV''
options at 30~TeV and 100~TeV. The result that a high energy muon
collider can attain on the size of the Higgs boson clearly exceed
the reach of other future collider projects, as reported by dashed
lines. 

Figure~\ref{fig:Lower-bounds-on-mStar} allows also to evaluate the
bounds for different amount of luminosity. In particular it highlights
that the luminosity requirement of eq.(\ref{eq:baseline-lumi}) is
quite close to the least possible luminosity necessary to meaningfully
run this analysis. In fact the solid black lines correspond to the
size of the deviation from the SM $Z\h$ total rate at which one has
to be sensitive to put a bound on $m_{\star}$ as strong as what indicated
by the shade of blue in the figure. The orange line for our baseline
luminosity runs parallel to these lines of iso-S/B and corresponds
to be sensitive to around 10\% deviations at 95\% CL. Thus if we imagine
to run a the same energy with lower luminosity we would be
effectively probing theories for which the EFT expansion parameter
eq.(\ref{eq:expansion-parameter}) has grown to be close to $O(1)$,
hence in a regime in which the EFT may not be valid. A simple way
to fall in this case is to reduce the mass of the new physics in eq.(\ref{eq:expansion-parameter}),
which eventually leads to $M<\s/2$ and thus makes the indirect search
strategy no longer meaningful. All in all, if we want to pursue indirect
new physics searches we need to push the energy of the machine, as
to profit from the growth with energy of the new physics effects,
but at the same time we need to keep a target luminosity around eq.(\ref{eq:baseline-lumi}),
or else the whole strategy of indirect new physics searches collapses.
In this eventuality the high energy muon collider would be a machine
suitable for direct new physics exploration, up to its kinematical
limit $\s/2$ for pair production, and with no meaningful sensitivity
whatsoever to new physics heavier than that.

A more complete analysis using other processes that involve Goldstone
and Higgs bosons has been carried out in Ref.~\cite{2012.11555v1},
which has analyzed $\h\h$,$\ww$,$\ww h$ as well as $Z\h$ production.
The results for the question on the size of the Higgs boson in some
of these processes depends on the strength on the interactions in
the BSM theory that completes the SM at the scale around $m_{\star}$,
therefore we give combined results in a plane $(m_{\star},g_{\star})$
in Figure~\ref{fig:Bounds-Mstar}. Also in this more refined setting
it is clear that the high energy muon collider options in the multi-TeV
regime can improve by orders of magnitude our knowledge on the point-like
nature of the Higgs boson. Thus a high energy muon collider operating
at 10 TeV can be said to be a magnifying glass a factor above 10 more
powerful than even the most powerful traditional colliders in discussion
in the future collider landscape. 

For completeness we remark that other bounds on the same plane can
be put if the top quark is also a composite particle with a finite
size, as studied in Ref.~\cite{2010.05915v1,4topsMuon}. These bounds
give even more support to the high energy muon collider as a most
powerful tool to study the Higgs boson and top quark nature as elementary
particles.

\subsection{Copious production of SM states}

The great fluxes of effective SM gauge bosons radiated off the beams
implies that SM final states with invariant mass $\sqrt{\hat{s}}\ll\s$
can be produced very abundantly. Very interestingly, these processes
have cross-section that grows logarithmically in this regime 
\begin{equation}
\sigma(VBF\to SM)\simeq O(1)\text{ pb}\cdot\log\left(\frac{\s/\text{TeV}}{\shat/0.1\text{TeV}}\right)\,,\label{eq:vbf-rate}
\end{equation}
where the O(1) factor accounts for the different values of the fluxes
of the type of boson considered in the fusion. Of course, when multiple
boson fusion channels are available this estimate must be adjusted,
e.g. $WW\to h$ , depending on the type of analysis one has in mind,
might be augmented by $ZZ,Z\gamma,\gamma\gamma\to h$ if one is not
tracking the presence of forward muons in the computation of a total
Higgs boson rate. Similarly, the production of colored particles or
particles with spin can change the multiplicity of final states and
the total rate will reflect the increased multiplicity of states.

Despite eq.(\ref{eq:vbf-rate}) is only a rough estimate of the rate
of producing relatively light SM states, it helps greatly to understand
the potential of a high energy muon collider in the search of new
physics over the so-called intensity frontier. 

\subsubsection{A Giga-Higgs boson program}

Following a luminosity scaling from the baseline eq.(\ref{eq:baseline-lumi})
we can anticipate a total production of Higgs bosons in the ballpark
of a fraction of a billion, e.g assuming 100~$\text{ab}^{-1}$ at
a 30 TeV collider and $\sigma_{h}\simeq1.2\text{ pb}$.

Such large number of Higgs bosons produced at a high energy muon collider
qualifies the machine as a Higgs boson factory. Indeed it is expected
to produce 100 times the number of Higgs bosons considered for the
most advanced low energy ``Higgs factories'', such as CEPC or FCC-ee
operating at $\s=240\text{ GeV}$. 

The large number of Higgs bosons expected at the high energy muon
collider will enable studies of the Higgs boson branching ratios in
rare decay modes with unprecedented precision, e.g. $h\to\mu\mu$,
$h\to\gamma\gamma$ and $h\to\gamma Z$ could be measured at, or even
below, the $1\%$ precision level. Being rare decay modes new physics
can be more visible in these channels.

Furthermore, new exotic rare decay modes of the Higgs boson can be
searched for with a potential of being sensitive to ultra-rare decay
modes down to $BR\simeq10^{-7}$. 

\begin{figure}
\includegraphics[width=0.49\linewidth]{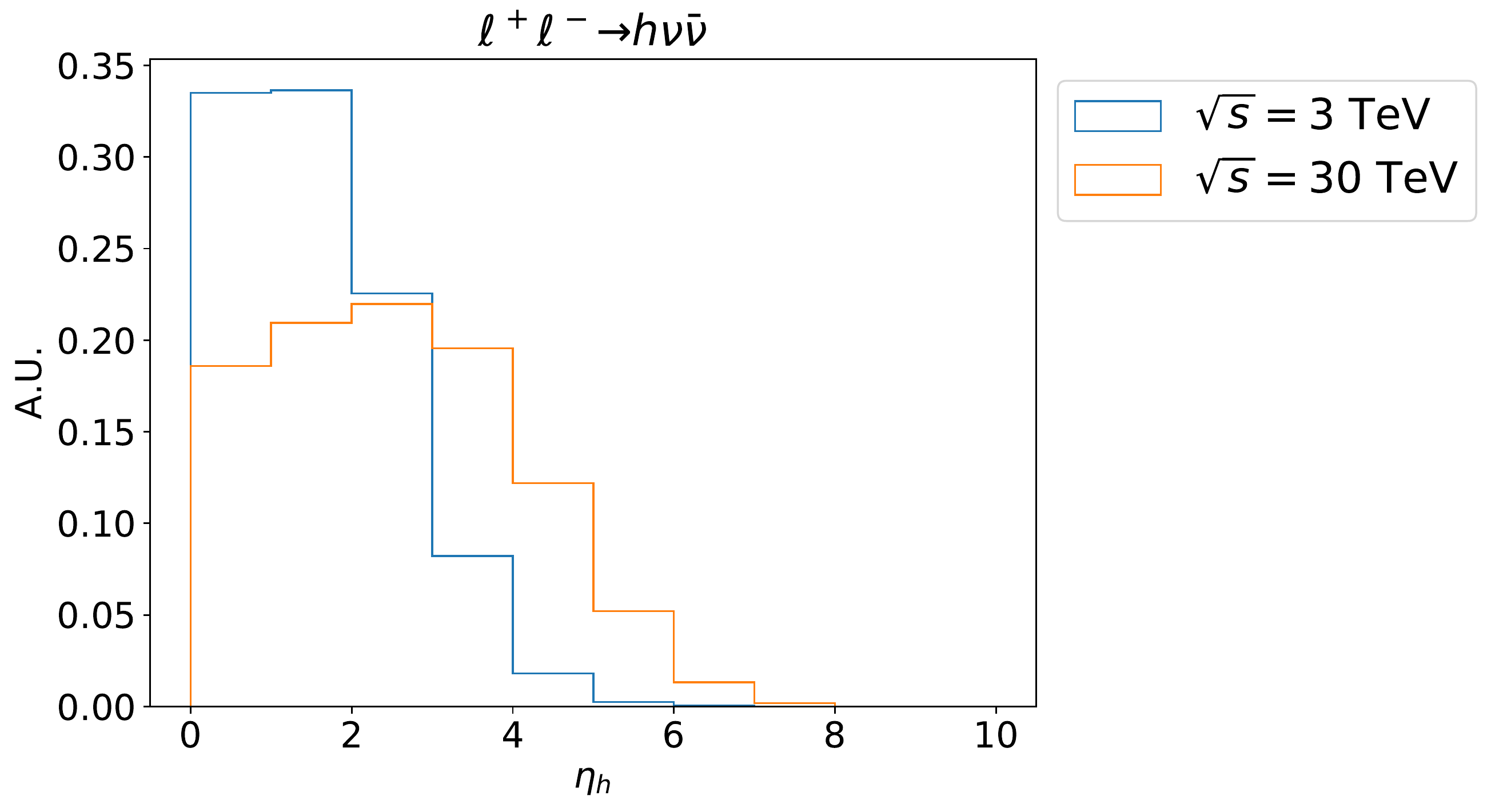}\includegraphics[width=0.49\linewidth]{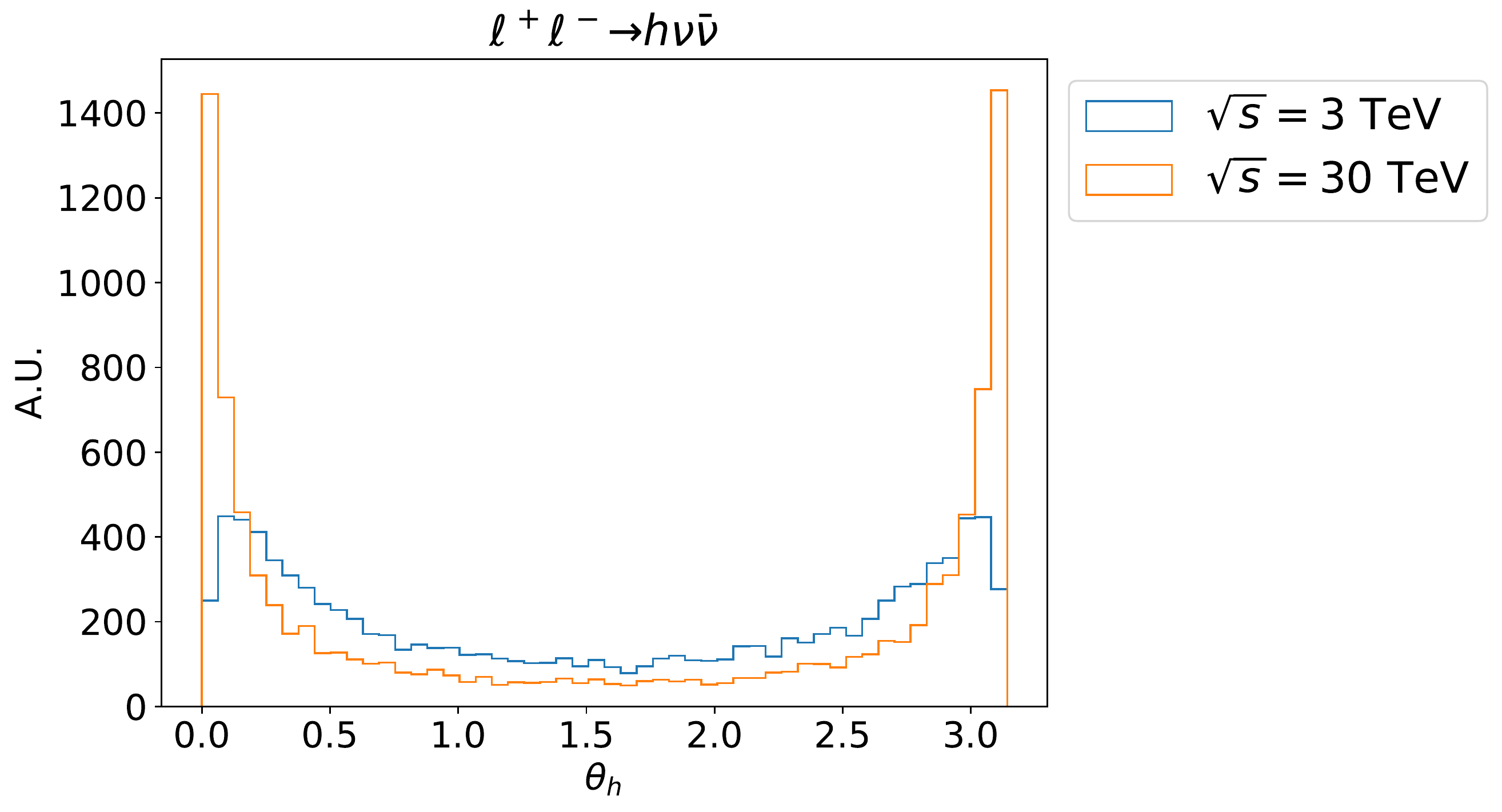}\\
\includegraphics[width=0.49\linewidth]{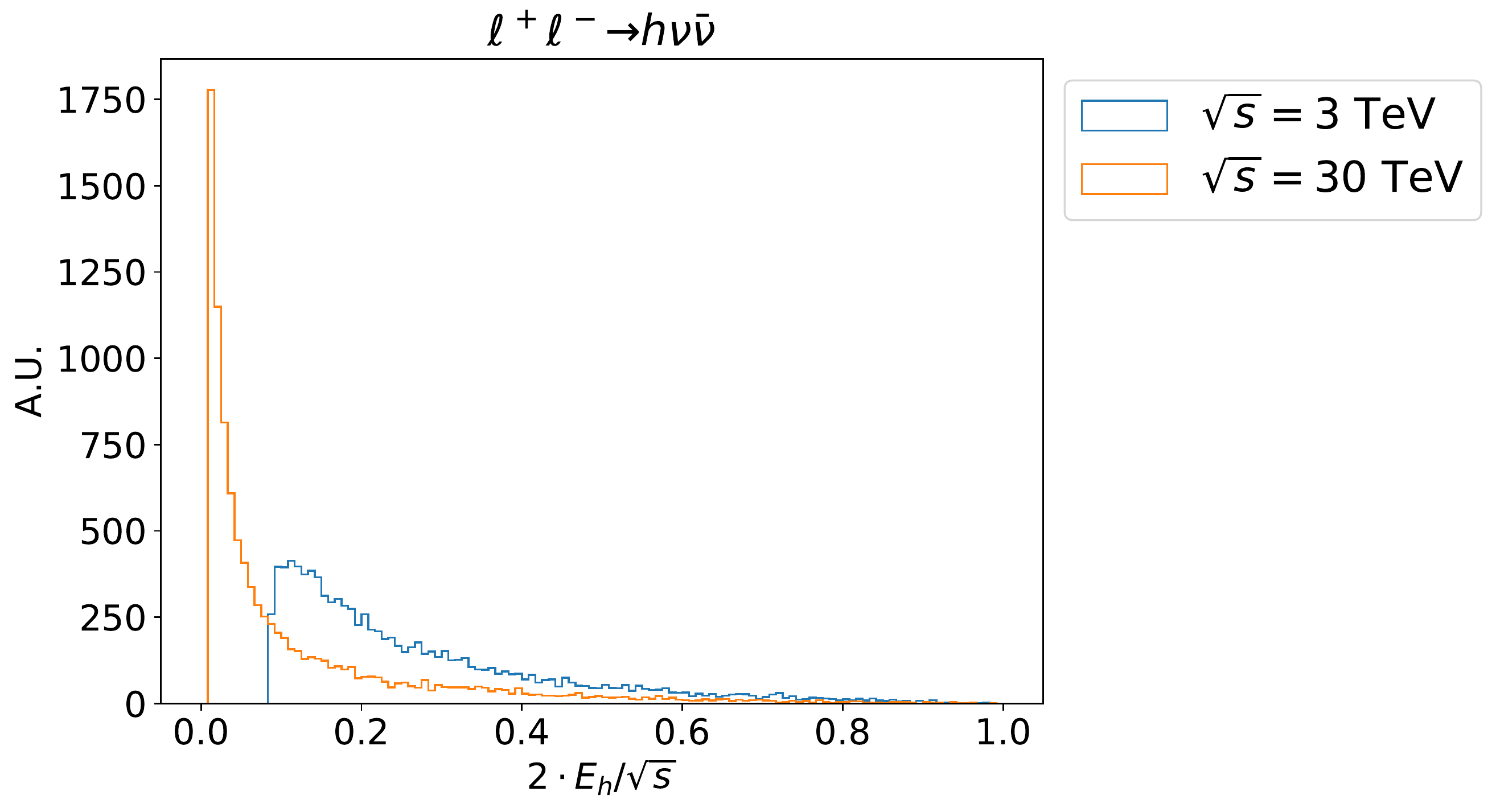}

\caption{\label{fig:Higgs-boson-direction}Higgs boson direction (as angle
$\theta$ or pseudo-rapidity $\eta$) and energy distributions in
the laboratory frame for $\llll\to\nu\bar{\nu}h$.}
\end{figure}

Of course, in order to achieve these results, it will be key to have
sufficiently hermetic detectors or put in place suitable detectors
dedicated to this kind of physics. In Figure~\ref{fig:Higgs-boson-direction}
we can observe how going towards higher energies the bulk of the Higgs
boson production tends to shift towards the beam pipe. Indeed, at
a 30 TeV muon collider roughly half the Higgs bosons would be produced
at large pseudo-rapidity $\eta_{h}>2.5$. Efforts have already started~\cite{2001.04431v1}
to study the detector performances for a moderately high energy muon
collider in the few TeV ballpark. Continuing work \cite{muonColldierDP}
on detector performances under the International Muon Collider Design
Study~\cite{MuonColliderCollaboration} has shown these early encouraging
results can be further improved. 
Phenomenological studies~\cite{2008.12204v1} have concluded that it is possible to measure the $hVV$ couplings using signatures with one Higgs boson plus unobserved forward beam remnants, e.g. neutrinos from the WBF Higgs production. Judiciously requiring the presence of a forward muon in the detector acceptance, the combination of the measured rates with and without this requirement allows the disentangled extraction of the $hZZ$ and $hWW$ couplings. 

Extraction of the triple and quadruple Higgs boson couplings as well as the gauge-Higgs quartic HHVV  have been studied in Refs.~\cite{2008.12204v1,2003.13628v1}. Remarkably, the trilinear Higgs coupling can be extracted with a precion around few percent and the $hhWW$ coupling with precision around $10^{-3}$, while the four $h$ coupling can be extracted with a precision around 50\%,  if the $hhh$ couplings is assumed to be as  predicted in the SM.

Considering only SM final states, a global analysis of the 
Higgs couplings extraction from the abundant production in VBF at a multi-TeV muon collider \cite{2103.14043v1} has shown that the large number of Higgs bosons produced can lead to a sub-permil determination of the $hWW$ coupling and to percent or sub-percent precision on the other couplings, including couplings involved in rare loop decay modes such as the $hZ\gamma$ coupling, and the bottom quark and muon Yukawa coupling.

\subsubsection{A Mega-Top quark program}

Besides being a Higgs boson factory, a high energy muon collider can
be - at the same time and under the same machine operating conditions
- a very effective top quark factory as well.

\paragraph{Low-energy top quarks}

The inclusive production of top quarks is dominated by associated
production with a pair of neutrinos which yields low-boost top quarks
pairs. The total cross-section is about constant $20-30\text{fb}$
from 3 to 30 TeV, and is clearly a lot lower than a low-energy lepton
collider, where fraction of $\text{pb}$ can be attained. However,
thanks to the luminosity eq.(\ref{eq:baseline-lumi}) expected at
a high energy muon collider the number of top quarks produced by $W$
fusion can be comparable, and even larger, than what can be attained
at machines operating around the threshold for the Drell-Yan production
such as $\ee$ machines at proposed $350$, $380$ or $500\,\text{GeV}$
dedicated stages. 

\begin{table}
{ 
\scriptsize
\begin{centering}
\begin{tabular}{|c|c|c|c|}
\hline 
$\sqrt{s}$ & $\sigma(\llll\to t\bar{t})$ & $\mathcal{L}$ & $\sigma\cdot\mathcal{L}$\tabularnewline
\hline 
\hline 
0.5 TeV & 548 fb & 4/ab & $2.2\text{M}$\tabularnewline
\hline 
3 TeV & 19 fb & $2.5$/ab & $47\text{K}$\tabularnewline
\hline 
30 TeV & 0.19 fb & 90/ab & $17\text{K}$\tabularnewline
\hline 
\end{tabular}~~%
\begin{tabular}{|c|c|c|c|}
\hline 
$\sqrt{s}$ & $\sigma(\llll\to\nu\nu t\bar{t})$ & $\mathcal{L}$ & $\sigma\cdot\mathcal{L}$\tabularnewline
\hline 
\hline 
0.5 TeV & 0.23 fb & 4/ab & $0.9\text{K}$\tabularnewline
\hline 
3 TeV & 5.4 fb & $5$/ab & $27\text{K}$\tabularnewline
\hline 
30 TeV & 31 fb & 90/ab & $2.7\text{M}$\tabularnewline
\hline 
\end{tabular}
\par\end{centering}
}
\caption{\label{tab:Top-quark-production} Top quark production cross-section
in Drell-Yan (left) and $W$ boson fusion (right) at $\s=0.5\text{ TeV}$,
$3\text{ TeV}$, $30\text{ TeV}$. These numbers are obtained at LO
in perturbation theory using $\mgfive$\cite{Alwall:2014bq}. The
luminosities used are those following eq.(\ref{eq:baseline-lumi})
for 30 TeV, whereas we use projected luminosities for top quark factory
operation of ILC at 0.5 TeV \cite{Barklow:2015aa} and CLIC 3 TeV\cite{Robson:2018ab}.
Radiative corrections and beam energy spreads should be taken into
account in a realistic setup but are not expected to change the overall
picture (e.g. at 3 TeV $\sigma(\llll\to t\bar{t})=25\text{fb}$ if
radiative corrections are included \cite{Abramowicz:2018aa}).}
\end{table}

A comparison of cross-sections and total number of top quarks produced
is reported in Table~\ref{tab:Top-quark-production}. The large flux
of partons that can produce top quark pairs is clearly sufficient
to pursue a full fledge program ``at the pole'' of the top quark
with similar measurements as those considered for other lepton colliders
with dedicated top quark physics stages~\cite{Abramowicz:2018aa,Zarnecki:2016lr,Janot:2015zp,Janot:2015dp}. 

\paragraph{High-energy top quarks}

The increase of luminosity with energy we have assumed in eq.(\ref{eq:baseline-lumi})
guarantees an approximatively constant number of top quarks produced
from large momentum transfer processes such as $2\to2$ scattering
$\llll\to t\bar{t}$. Therefore it is possible to carry out measurements
at large momentum transfer keeping statistical uncertainty constant
even if the collider energy considered varies. With such provision
we can quickly estimate the reach of a high energy muon collider in
a similar way to what we have seen for diboson processes. Following
the diboson path~\cite{2012.11555v1} we can put very stringent bounds
on contact interactions that involve top quarks and give rise to effects
that grow with momentum transfer. In this aspect a high energy muon
collider has an increased potential to probe new physics at high energy
and it essentially offers the best reach for a collider of same beam-beam
center of mass energy. Preliminary results presented in Ref.~\cite{4topsMuon}
confirm these estimates, but more refined studies are needed.

While a dedicated study of the reach for new physics using high-momentum
transfer $\llll\to t\bar{t}$ production is not yet available, we
can quickly estimate the expected performances extrapolating from
CLIC 3 TeV studies. Beamstrahlung and ISR effects are different for
a muon collider and a $\ee$ linear collider, still we can obtain
a reliable estimate of the ballpark of the reach of a simple angular
distribution study of new physics effects in DY production. A 30 TeV
muon collider can be sensitive to new physics from mass scales well
in excess of 100~TeV. 

\paragraph{Further production modes and measurements: $e^{+}e^{-}\to t\bar{t}h+X$
the Yukawa coupling $y_{t}$ and more }

Exploiting the large center of mass energy it is possible to produce
richer final states than the simple $t\bar{t}$. For instance it is
possible to obtain the $t\bar{t}h$ final state, that is sensitive
to the top quark Yukawa coupling and is characterized by large momentum
transfer. 

This process can give a direct measurement of $y_{t}$ with a precision
around few percent. While the precision on $y_{t}$ per se is not
much different from what can be obtained at less energetic colliders,
e.g. the 3 TeV stage of CLIC, the fact that this measurement is characterized
by a much larger momentum transfer makes it much more sensitive to
possible new physics contributions. A through study of the sensitivity
of $t\bar{t}\h$ to new physics in a clean EFT language is still missing,
however we can expect that this measurement will be very sensitive
to contact interactions, thanks to the benefit from running at high
energy. 

At the high energy muon collider it will be possible to produce top
quarks and Higgs bosons in even more complex final states such as
$t\bar{t}h\nu\bar{\nu}$. Also this process is sensitive to the Yukawa
coupling of the top quark, but is characterized by the typical momentum
transfer of the WBF processes, hence it is not going to be a most
powerful probe of contact interactions involving the Higgs boson and
the top quark. Putting this process together with the high momentum
transfer $t\bar{t}\h$ we expect to be able to constrain both new
physics effects that are magnified by the momentum transfer, hence
are suppressed by an EFT expansion parameter similar to eq.(\ref{eq:expansion-parameter}),
and those that are insensitive to the momentum transfer, e.g. those
which can be cast as pure shifts of SM parameters. 

\begin{table}
\begin{centering}
\begin{tabular}{|c|c|c|c|c|c|}
\hline 
$\s$ & $\sigma(\llll\to t\bar{t}h)$ & $\mathcal{L}$ & $\sigma\cdot\mathcal{L}$ & $\frac{\delta\sigma}{\sigma}$ at 68\% CL & $\frac{\delta y_{t}}{y_{t}}$ at 68\% CL\tabularnewline
\hline 
\hline 
30 TeV & 7 ab & 90/ab & 630 & 4.0\% & 2.0\%\tabularnewline
\hline 
 & $\sigma(\llll\to t\bar{t}h\nu\nu)$ & $\mathcal{L}$ & $\sigma\cdot\mathcal{L}$ & $\frac{\delta\sigma}{\sigma}$ at 68\% CL & $\frac{\delta y_{t}}{y_{t}}$ at 68\% CL\tabularnewline
\hline 
30 TeV & 100 ab & 90/ab & $9000$ & 1\% & 0.5\%\tabularnewline
\hline 
\end{tabular}
\par\end{centering}
\caption{\label{tab:Expected-rates-for-tth}Expected rates for $t\bar{t}\h+X$
reactions and an estimate on the sensitivity to energy-independent
effects, such as a shift in the Yukawa coupling of the top quark.}
\end{table}

For a quick estimate of the power of these processes in constraining
new physics we can look at 30 TeV muon collider rates reported in
Table~\ref{tab:Expected-rates-for-tth} and the resulting statistical
uncertainties on these rates. Keeping in mind that high momentum transfer
$t\bar{t}\h$ rates should be used to put bounds on effects of new
physics that grow with energy, we can see how at a 30 TeV machine
the two classes of processes have a similar statistical uncertainty,
hence at this large center of mass energies it is possible to simultaneously
carry out new physics searches and put meaningful bounds using both
types of processes.

All in all it is possible to imagine a rich physics case for the study
of new physics involving the top quark and the Higgs boson at the
high energy muon collider. Learning the results sketched above a larger
set of processes can be imagined for an extended program on the top
and bottom quarks and Higgs boson sector involving $b\bar{b}+X,b\bar{b}\h+X$,
$tb+X,$ and $tb\h+X$ processes. Preliminary results on this enlarged
set of processes~\cite{RFinprep} indicate that they have a constraining
power similar to di-boson processes on new physics scenarios where
the Higgs boson and the third generation quarks are not elementary
point-like states.

\section{Conclusions\label{conclusions}}

The Higgs boson is a cornerstone of the Standard Model of particle physics, as it provides   a concrete realization of the mechanism of spontaneous symmetry breaking needed to separate electromagnetic and weak gauge interactions. The Higgs boson is also a unique and singular object in the present formulation of the SM. In fact it is the only Lorentz scalar in the model and needs to be exactly point-like for the model to be consistent. At the same time, the Higgs boson mass and its properties have a remarkable sensitivity to the existence of new heavy states, whose mass acts as a source of destabilization of the weak scale.  With such unique role in the SM and special properties in QFT in general, the Higgs boson is a most important target for studies to be carried out at future particle physics facilities. 

In this contribution we have highlighted the possible studies that a low energy muon collider might enable to better understand the nature of the Higgs boson. We have outlined significant challenges for the use of data coming from Higgs bosons produced from resonant annihilation of muons beams. The quality of the beam, and in particular its energy spread, turns out to be a key parameter to assess the outcome of Higgs boson factory at the pole. If a machine at the Higgs boson pole could be realized with relative beam energy spread  O( $3\cdot 10^{-5}$), and a few fb$^{-1}$ integrated luminosity accumulated, the results on the Higgs boson couplings would bring a significant improvement over the most optimistic HL-LHC projections. In the landscape of future colliders and their performance on the determination of Higgs boson properties these improvements generally fall short compared to other projects. However, it should be remarked that a Higgs boson factory from resonant muon annihilation might provide the best measurement of the Higgs boson coupling to muons and might be one of the few ways, if not the only one,  to directly measure the Higgs boson width with good precision. 

In the second part of our contribution we have discussed the possibility of using a high energy muon collider to study the Higgs boson and in general the Higgs sector and the physics  BSM associated to it. We have outlined a physics program that can be pursued at a multi-TeV muon collider by leveraging both high rate reactions at low momentum transfer, such as the vector boson fusion production of Higgs bosons and other SM states, and the high momentum transfer reactions such as direct Drell-Yan annihilation processes into SM states or possible BSM final states. Concerning the physics of the Higgs boson we have highlighted the possibility to study contact interactions involving the Higgs boson or longitudinal gauge bosons (or both) as a mean to study new physics in the Higgs sector. We have discussed how this search for new physics effects demands the operation of such a multi-TeV machine with sufficient luminosity to be able to study at the few percent level the total rate of the least abundant SM Drell-Yan process, e.g. $\mumu \to Z\h$. 

A machine designed to collect around 10~ab$^{-1}$ at 10~TeV center of mass energy can potentially probe new physics related to the breaking of the electroweak symmetry and the Higgs boson up to mass scales just short of 100 TeV. 

The luminosity requirement outlined above in eq.(\ref{eq:baseline-lumi}) for the investigation of new physics related to the Higgs boson would enable also a host of investigation for contact interactions of SM states that can be generated by new physics. Therefore the achievement of these luminosity targets would put experiments run at the high energy muon collider in position to be sensitive to a large number of new physics scenarios. Furthermore, the collection of such large luminosity would enable the precision study of SM states produced in low momentum transfer reactions from a data-set of unprecedented size  and the unique feature of being produced from purely electro-weak reactions. These studies of SM states would complement beautifully with the study of contact interactions from new physics, essentially ``multiplexing'' the physics case of the muon collider. 

Should the luminosity requirement indicated above not be met, the high energy muon collider remains a fantastic machine to explore the energy frontier. In fact, it provides a clean environment to study the results of high energy collisions and at the same time can probe very large mass scale, thus putting together the best of the $\ee$ and $pp$ colliders features.  The direct search of new heavy particles is a key ability of a high energy muon collider, as it can probe heavy new physics charged under electro-weak gauge interactions, which is ubiquitous in new physics models. Thanks to the clean collision environment a high energy muon collider operating even 2 orders of magnitude below the luminosity requirement discussed for indirect new physics searches would be able to discover new particles up to about $\s/2$, hence swiping the whole range from the HL-LHC limits to the multi-TeV mass range.

Although the direct search of new physics states is an exciting and potentially rewarding program, it is very important to stress that designs aimed at the  luminosity requirement outlined for indirect searches of new physics may be even more rewarding and far reaching. The consequence of establishing the feasibility of a baseline luminosity $$\mathcal{L}=10\, {\rm ab}^{-1} \left( \frac{\s}{10\, {\rm TeV}}\right)^2\,,$$ would be momentous. In fact, by going at higher energies while increasing luminosity it would lead to a path of systematically improving the results described above testing   new physics mass scale that grow linearly with center of mass energy. 

It is important to stress that this possibility is unique to muon colliders. In fact, at variance with circulating 
and linear
 electron and positron beams, muon beams can be manipulated so that it is in principle possible to 
 reach a luminosity per unit wall plug power that grows as the beam energy grows~\cite{Delahaye:2019aa,1808.01858v2}.
Therefore, muon beams allow to entertain the idea of collisions at  even higher center of mass energies in the tens of TeV. Thanks to the relatively low power cost ``per TeV'' center of mass energy of these machines we can reasonably imagine to extend the high energy muon collider physics program at higher energies with instantaneous luminosity that grow as $s$, thus keeping a fixed amount of recorded events for the simplest Drell-Yan annihilations. 
Along this line we can imagine an upgrade path for the investigations of new physics related to the Higgs boson that for a center of mass energy of 30 TeV would be probing mass scales of new physics in the range of hundreds of TeV.

\section*{Acknowledgements}
It is a pleasure to thank Zhen Liu for discussions on the precision of the couplings determination at low energy. RF thanks Dario Buttazzo and Andrea Wulzer for many discussions and exchanges on the physics potential of the muon collider.

\bibliographystyle{JHEP}
\bibliography{bibliographyfile}

\end{document}